\documentclass[conference]{IEEEtran}
\usepackage{cite}            % auto-sort and range-compress [1],[2],[5]--[7]

\ifCLASSINFOpdf
  \usepackage[pdftex]{graphicx}
  \graphicspath{{images/}}
  \DeclareGraphicsExtensions{.pdf,.png,.jpg}
\else
  \usepackage[dvips]{graphicx}
  \graphicspath{{images/}}
  \DeclareGraphicsExtensions{.eps}
\fi

\usepackage{amsmath}
\usepackage{amssymb}
\usepackage{amsfonts}
\usepackage{array}
\usepackage{booktabs}        % \toprule \midrule \bottomrule
\usepackage{multirow}        % row-group labels spanning multiple rows
\usepackage{stfloats}        % allow table* (wide floats) at bottom of 2-col pages

\usepackage[caption=false,font=footnotesize]{subfig}

\usepackage[table]{xcolor}
\definecolor{controlrow}{gray}{0.92}  % light gray for in-distribution control rows
\usepackage{url}
\usepackage{textcomp}

\newcommand{\up}{$\uparrow$}               % higher-is-better
\newcommand{\down}{$\downarrow$}           % lower-is-better
\newcommand{\toone}{$\rightarrow\!1$}      % target value 1.0

\begin{document}

% ----------------------------------------------------------------------------
% TITLE  (no math/special symbols in title per IEEEtran)
% ----------------------------------------------------------------------------
\title{Physics-Preserving Latent Compression for Zero-Shot Resolution Transfer in 3D Turbulence}

% ----------------------------------------------------------------------------
% AUTHORS — arXiv (public) version. De-anonymized. The ICDM submission
% version (main-ICDM.tex) keeps the anonymous block for triple-blind review.
% ----------------------------------------------------------------------------
\author{
Yilong Dai\textsuperscript{1},
Yiming Sun\textsuperscript{2},
Yiheng Chen\textsuperscript{1},
Ziyi Wang\textsuperscript{3},
Shengyu Chen\textsuperscript{2},
Xiaowei Jia\textsuperscript{2},
Runlong Yu\textsuperscript{1,\textdagger}
\\
\textsuperscript{1}The University of Alabama \quad
\textsuperscript{2}University of Pittsburgh \quad
\textsuperscript{3}University of Maryland, College Park
\\
{\small\texttt{\{ydai17, ychen226, ryu5\}@ua.edu,}}
{\small\texttt{\{yimingsun, shc160, xiaowei\}@pitt.edu, zoewang@umd.edu}}
\\
\textsuperscript{\textdagger}Corresponding author.
}

\maketitle

% ----------------------------------------------------------------------------
% ABSTRACT  (bullets-as-comments; prose filled in later. No math/cites here.)
% ----------------------------------------------------------------------------
\begin{abstract}
High-resolution turbulence modeling is essential for scientific computing, but remains constrained by the cost of direct numerical simulation and the scarcity of full-resolution data. Existing scientific compressors reduce storage but typically operate on per-frame representations, whereas learned compressors yield compact latents that are often resolution-dependent and weakly aligned with the physics of turbulence. This raises the need for a compression framework that reduces data size, preserves physical diagnostics, and transfers from low-resolution training fields to high-resolution test fields without retraining. In this paper, we propose \emph{Physics-Preserving Latent Compression (PPLC)}, a patch-local latent compressor for three-dimensional turbulence. Motivated by inertial-range scale similarity, PPLC treats fixed-size patches as transferable units and applies a shared variational autoencoder independently of the global grid size. It combines exact mean preservation, zero-mean fluctuation encoding, an invertible Haar wavelet front-end, shift-consistency regularization, and overlap-aware reconstruction. Instantiated on forced isotropic turbulence, PPLC is trained only on stride-downsampled $256^3$ fields and transfers zero-shot to $1024^3$ fields. Experiments show that PPLC improves the balance between reconstruction accuracy and physical fidelity over classical and learned baselines, keeping diagnostics such as dissipation, enstrophy, energy spectra, and incompressibility closer to the ground truth. Beyond turbulence compression, PPLC offers a general strategy for physics-preserving latent representations that support data-efficient scientific surrogate modeling.

\end{abstract}

% \noindent Code is available at \url{https://anonymous.4open.science/r/PPLC4ICDM-B751}.


\IEEEpeerreviewmaketitle

% ----------------------------------------------------------------------------
% BODY — each section is a separate file under sections/.
% ----------------------------------------------------------------------------

\section{Introduction}
\label{sec:intro}

Turbulent flows arise throughout science and engineering, but their chaotic and multiscale nature makes them difficult to simulate, store, and learn from. Direct numerical simulation (DNS) resolves the Navier--Stokes equations down to the smallest dynamically relevant scales, including the dissipative range. This physical fidelity requires a grid sufficiently fine to resolve the separation between the largest energetic motions and the smallest turbulent eddies. As the Reynolds number increases, this scale separation widens, forcing DNS to use increasingly large spatial grids and many time steps. High-resolution DNS is therefore expensive to generate, and the resulting data remain scarce.

This scarcity creates a basic tension for neural turbulence surrogates~\cite{li2021fno,kovachki2023neuraloperator,kochkov2021ml4cfd,stachenfeld2022learned}. These models are designed to emulate high-fidelity turbulent dynamics, but training them directly on full-resolution DNS data is often computationally prohibitive. A common workaround is to train on lower-resolution data obtained by subsampling high-resolution fields. This reduces the training cost, but it also removes small-scale structures that influence dissipation, enstrophy, spectral behavior, and incompressibility. The surrogate is then trained on data from which part of the target physics has already been filtered out. The central problem is therefore to construct a compact representation that reduces data size, preserves turbulence-relevant physical quantities, remains usable by downstream dynamics models, and transfers from low-resolution training fields to high-resolution test fields.

Compression offers a promising approach to addressing this problem, although existing methods are not fully aligned with these requirements. Learned compressors developed for natural images and videos~\cite{kingma2014vae,esser2021taming,rombach2022ldm,chen2025dcae,zheng2025rae,li2025wfvae,nvidia2025cosmos,hacohen2025ltxvideo} are usually optimized for perceptual quality or pixel-level reconstruction. Such objectives are effective for natural signals, but they do not directly control physical quantities that matter in turbulence, including the energy spectrum, dissipation, enstrophy, and the divergence-free constraint. Classical scientific compressors, such as proper orthogonal decomposition~\cite{lumley1967pod,sirovich1987snapshot}, wavelet thresholding~\cite{daubechies1988wavelets}, fixed-rate floating-point coding~\cite{lindstrom2014zfp}, error-bounded lossy compression~\cite{dicappello2016sz,ainsworth2018mgard}, and tensor decompositions~\cite{oseledets2011tt,schollwock2011dmrg}, provide useful tools for reducing simulation data. However, they are typically designed as per-frame storage representations. The latent spaces they produce are not generally learned as resolution-transferable state spaces on which neural dynamics models can operate directly. Existing learned compressors are also often tied to the resolution distribution observed during training, leaving open the question of whether a compressor trained only on lower-resolution fields can be applied, without retraining, to much higher-resolution turbulent data.

Our starting point is a physical observation from Kolmogorov's similarity theory~\cite{kolmogorov1941,pope2000turbulent}. In the inertial range of sufficiently developed turbulence, local velocity-increment statistics are governed primarily by scale and dissipation, rather than by the absolute grid on which the flow is sampled. This observation does not imply that patches at different resolutions are identical. It suggests that local turbulent structures, after accounting for scale, can have related statistical organization across resolutions when they lie in comparable inertial-range regimes. This motivates a patch-local compression principle: a fixed-size spatial patch can serve as the transferable unit, and a shared encoder trained on patches from one resolution can be applied to patches from another because its computation is independent of the global grid size.

Building on this principle, we propose \emph{\textbf{Physics-Preserving Latent Compression (PPLC)}}, a resolution-independent latent compressor for three-dimensional turbulence. PPLC partitions the field into fixed-size patches and compresses each patch with a shared convolutional variational autoencoder. This patch-local design makes the encoder--decoder independent of the global resolution: a larger field contains more patches, while each patch is processed by the same learned map. To preserve physical structure within each patch, PPLC separates the per-channel mean and stores it exactly, allowing the network to focus on the zero-mean turbulent fluctuation. The fluctuation is then passed through an invertible Haar wavelet transform, which reorganizes the signal by scale before learned encoding. To reduce artifacts introduced by patchwise processing, the model further uses a shift-equivariance consistency loss during training, together with overlap-aware reassembly at inference.

This design provides a principled route to zero-shot resolution transfer. In our setting, the compressor is trained only at $256^3$, using stride-downsampled fields that retain the large scales resolved at that resolution. At test time, the same encoder and decoder are applied unchanged to $1024^3$ fields, where each fixed-size patch contains finer-scale structures unseen during training. A fixed-size patch captures relatively large-scale structures at $256^3$ and relatively small-scale structures at $1024^3$; by inertial-range scale similarity, these patches can have related local statistics after accounting for scale. In this way, training remains inexpensive while reconstruction is evaluated in the high-resolution regime.

Beyond storage, the latent representation is meant to be acted on. The neural surrogates motivating this work advance turbulent states forward in time~\cite{dai2026flowrefiner,dai2026pest}, so a useful compressed representation should provide a state space on which such models can operate directly. As an initial test of this property, we include an in-distribution forecasting experiment showing that a model can roll the flow forward more accurately in the compressed latent space than in pixel space.

Our contributions are summarized as follows:
\begin{itemize}
    \item We propose a physics-motivated compression principle based on inertial-range scale similarity. This principle allows a compressor trained on low-resolution turbulent fields to be applied directly to high-resolution fields without retraining.
    
    \item We introduce PPLC, a patch-local latent compressor for three-dimensional turbulence. PPLC combines exact mean preservation, zero-mean fluctuation encoding, an invertible Haar wavelet front-end, shift-consistency regularization, and overlap-aware reconstruction to better retain both visual accuracy and turbulence-relevant physical quantities at high compression ratios.
    
    \item Experiments show that PPLC outperforms classical and learned compression baselines on both reconstruction and physical diagnostics, with an auxiliary forecasting test indicating that its latent space remains usable for downstream dynamics.
\end{itemize}

% Section 2 — Related Work. Run-in headings (no subsections), prose. All cites verified.
\section{Related Work}
\label{sec:related}

\textit{\textbf{Learned compression.}} Compressing data into a learned latent underpins modern generative modeling, from the variational autoencoder~\cite{kingma2014vae} and its disentangled~\cite{higgins2017betavae} and discrete~\cite{vanderoord2017vqvae} variants to the autoencoders behind latent diffusion~\cite{esser2021taming,rombach2022ldm}, with higher ratios via residual autoencoding~\cite{chen2025dcae} or pretrained-encoder reuse~\cite{zheng2025rae} and 3D video tokenizers~\cite{li2025wfvae,nvidia2025cosmos,hacohen2025ltxvideo}. Such latents increasingly serve as a state space for downstream dynamics, as in latent world models~\cite{dong2026worldmodel}. All optimize perceptual or pixel reconstruction, however, leaving the small-scale structure and conservation laws of turbulence unconstrained.

\textit{\textbf{Scientific data compression.}} A parallel line respects the physics of simulation data. Transform methods truncate a basis, including proper orthogonal decomposition~\cite{lumley1967pod,sirovich1987snapshot}, wavelet thresholding~\cite{daubechies1988wavelets}, and fixed-rate floating-point coding~\cite{lindstrom2014zfp}, while tensor-train factorizations~\cite{oseledets2011tt}, equivalent to matrix product states~\cite{schollwock2011dmrg}, exploit low rank; error-bounded compressors such as SZ~\cite{dicappello2016sz} and MGARD~\cite{ainsworth2018mgard} guarantee pointwise error. These are physically grounded but tied to a single resolution and per snapshot, with no notion of temporal evolution and not meant to be acted on downstream.

\textit{\textbf{Function-space models.}} Closer to resolution transfer are function-space models. Neural operators learn discretization-invariant maps between function spaces~\cite{li2021fno,kovachki2023neuraloperator,lu2021deeponet}, some with explicitly multi-scale structure via U-shaped~\cite{rahman2023uno} or multiwavelet~\cite{gupta2021multiwavelet} kernels, an idea our Haar front-end shares. Functional autoencoders~\cite{bunker2025fae} and coordinate-based fields~\cite{sitzmann2020siren,perez2018film,guo2025cnf} are resolution-invariant but typically fit per snapshot or solve a fixed PDE~\cite{dai2026pdesurvey}; we instead keep a convolutional, patch-local bias with a single shared encoder and target compression rather than operator learning.

\textit{\textbf{Machine learning for turbulence.}} Our approach rests on the inertial-range scale self-similarity of turbulence~\cite{kolmogorov1941,pope2000turbulent}. Machine learning has accelerated fluid simulation~\cite{kochkov2021ml4cfd,raissi2019pinn,stachenfeld2022learned}, and image generative backbones~\cite{ho2020ddpm,peebles2023dit} have been adapted to reconstruct, super-resolve~\cite{shu2023diffusion}, and forecast 3D turbulence on JHTDB~\cite{dai2026flowrefiner,dai2026pest}. These operate on full-resolution fields; our work is upstream of them, providing a compact physics-preserving latent in which such forecasting can be carried out.

% We evaluate on the Johns Hopkins Turbulence Database~\cite{perlman2007jhtdb,wan2016jhtdb}.

% Section 3 — Method. patch 32^3, latent C x 8^3, S_k uses k=8.
\section{Method}
\label{sec:method}

% ---- Figure 1: method/architecture overview (double-column float) ----------
% Figure source/prompt: ../figures_source/figure1_prompt.txt (notes: figure1_notes_ZH.md)
\begin{figure*}[t]
  \centering
  \includegraphics[width=\textwidth]{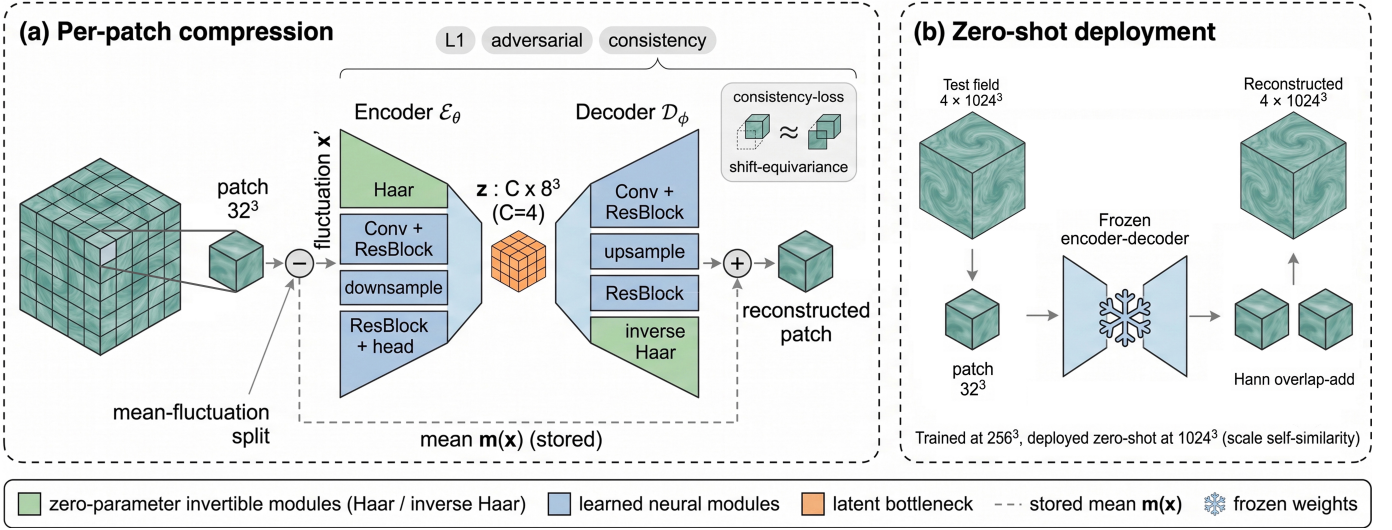}
  \caption{\textbf{Method overview.} (a) Per-patch compression and (b) zero-shot
  deployment at a higher resolution.}
  \label{fig:overview}
\end{figure*}

\subsection{Problem Setup}
\label{subsec:setup}
We consider a turbulent velocity--pressure field $X \in \mathbb{R}^{4 \times N \times N \times N}$, with the three velocity components and pressure stacked as channels, sampled from incompressible isotropic turbulence. Our goal is to compress such a field at a high ratio while meeting three requirements at once: the reconstruction should be \emph{accurate} in a pixel sense, it should be \emph{physically faithful}, keeping the energy spectrum, dissipation, enstrophy, and the incompressibility (divergence-free) constraint as close to the ground truth as possible, and its latent representation should be \emph{usable} by a downstream model that simulates the flow, rather than merely decodable. We additionally target a property that fixed-resolution compressors do not provide: \emph{resolution transfer}. Concretely, we train at a single resolution and apply the learned encoder and decoder, unchanged, to fields of a much higher resolution at test time. Figure~\ref{fig:overview} gives an overview of the per-patch compression and the zero-shot deployment.

This last requirement shapes the entire design: a model whose behavior depends on the absolute grid size cannot transfer across resolutions. The central question is therefore how to build an encoder--decoder pair whose computation is independent of the global resolution, yet still captures the local physical structure of the flow. We develop such a design below, starting from the physical principle that makes resolution transfer possible.

\subsection{Physical Principle: Scale Self-Similarity}
\label{subsec:principle}
Our design rests on Kolmogorov's similarity theory~\cite{kolmogorov1941,pope2000turbulent}, the physical principle that makes resolution transfer possible. In the inertial range of high-Reynolds isotropic turbulence, the second-order velocity structure function follows the scaling
\begin{equation}
  S_2(r) = \big\langle \lVert u(x{+}r) - u(x) \rVert^2 \big\rangle \;\sim\; (\varepsilon\, r)^{2/3},
  \label{eq:k41}
\end{equation}
for a separation $r$ within the inertial range, where $\varepsilon$ is the mean dissipation rate. The statistics of velocity increments thus depend only on the scale $r$ and on $\varepsilon$, not on the absolute grid. This is the key observation we exploit: a fixed-size spatial window placed on a coarse grid spans a band of \emph{larger} physical scales than the same window on a fine grid, but as long as both bands lie in the inertial range, Eq.~\eqref{eq:k41} makes their internal statistics similar up to the factor $\varepsilon$. A model that learns to compress such a window at one resolution should therefore apply, statistically, at another. Two properties of the design below turn this statistical similarity into actual transfer: the encoder reads only a $32^3$ neighborhood through translation-equivariant convolutions, so it depends on local content rather than the global grid, and the patch-wise reconstruction makes global quality a function of per-patch quality, which the scaling above holds fixed across resolutions.

\subsection{Patch-Local Compression}
\label{subsec:patch}
The key to resolution independence is to keep the model from ever seeing the full field. Instead of encoding $X$ as a whole, we partition it into non-overlapping cubic patches of fixed size $P=32$,
\begin{equation}
  X = \bigcup_i x_i, \qquad x_i \in \mathbb{R}^{4 \times P \times P \times P},
  \label{eq:patch}
\end{equation}
and compress every patch independently with the same encoder. Because the encoder operates only on a $32^3$ window, its parameters and computation are tied to the patch, not to $N$: a field of any resolution is simply a different number of patches. A $256^3$ field yields $512$ patches and a $1024^3$ field yields $32{,}768$, but each is processed identically. Resolution transfer thus reduces to a single question, namely whether a patch learned at one resolution is representative of a patch at another, which the scale self-similarity above answers.

This choice also lets us train economically. We obtain training data by downsampling $1024^3$ fields to $256^3$ with a stride of four. Such global downsampling removes the smallest scales of the full field, but a fixed-size patch on the coarse grid simply covers a band of larger scales than the same patch would on the fine grid. By the scale self-similarity of inertial-range turbulence, these larger scales are statistically similar to the smaller ones the patch will see at test time, so the patch decomposition turns a property of turbulence into a practical training strategy.

\subsection{Mean--Fluctuation Decomposition}
\label{subsec:cms}
Within a single patch, not everything needs to be learned. A patch carries a per-channel mean, its bulk velocity and pressure offset, that is cheap to store exactly but, if left in place, would force the network to spend capacity reproducing a constant. We therefore apply a per-patch mean--fluctuation decomposition (a DC/AC split), separating each patch channel-wise into its spatial mean and a zero-mean fluctuation,
\begin{equation}
  m(x) = \frac{1}{P^3}\sum_{\text{voxels}} x \;\in\; \mathbb{R}^{4},
  \qquad
  x' = x - m(x),
  \label{eq:cms}
\end{equation}
storing the four mean values $m(x)$ verbatim and passing only the fluctuation $x'$ to the encoder. Storing the mean exactly is lossless and lets the learned capacity focus on the fluctuating structure, which is where the turbulent physics lives. The encoder is thus a model of fluctuations, not of absolute levels.

\subsection{Encoder}
\label{subsec:encoder}
The fluctuation $x'$ still mixes all spatial frequencies, and a turbulent patch concentrates very different information at different scales. Before any learning, we therefore apply a single-level 3D Haar wavelet transform, which maps the $32^3$ patch to a $16^3$ tensor with eight frequency sub-bands stacked along the channel axis. The transform is exactly invertible and has no parameters, so it loses nothing and adds no cost; it simply decorrelates the frequency bands so that the subsequent convolutions operate on an already-structured input. From there the encoder is a 3D convolutional stack that lifts the channels, applies residual blocks, and performs a single strided downsampling to an $8^3$ grid, ending in a variational head that outputs the latent posterior:
\begin{equation}
  z = E_\theta(x'), \qquad z \in \mathbb{R}^{C \times 8 \times 8 \times 8}.
  \label{eq:encoder}
\end{equation}
We keep a single learned downsampling step on purpose. Downsampling is where small-scale information is most easily lost, so minimizing its use, and pairing it with the lossless Haar front-end, limits the irreversible mixing that erodes the spectrum.

A deliberate design choice is the \emph{shape} of $z$. Rather than collapsing the patch into a few channels over a tiny $4^3$ grid, we retain an $8^3$ spatial layout with a small channel count $C$. The compression ratio is set by $C$: at $C=4$ the latent holds $4\cdot 8^3$ values per patch, a $64\times$ reduction once the four mean values are included. Preserving spatial extent in the latent, rather than folding it into channels, keeps neighborhood structure that the downstream simulator relies on and that a channel-heavy latent would discard.

\subsection{Decoder and Reconstruction}
\label{subsec:decoder}
The decoder mirrors the encoder: it lifts the latent, applies residual blocks, performs a single transposed-convolution upsampling back to $16^3$, and an inverse Haar transform returns the patch to $32^3$ with its original four channels. The mean removed earlier is then added back, giving the reconstructed patch
\begin{equation}
  \hat{x} = m(x) + D_\phi(z).
  \label{eq:decoder}
\end{equation}
The decoded patches are then reassembled into the full field. Tiling them on the original non-overlapping grid is the simplest choice, but it makes the patch boundaries a source of discontinuity: a map that is not perfectly consistent across a seam leaves a periodic artifact at multiples of the patch period in the energy spectrum. We address this at two levels. At training time, the consistency term in the objective below makes the per-patch map insensitive to where a boundary falls, attacking the discontinuity at its source. At test time, we further reassemble with a Hann-windowed overlap-add: patches are decoded on a half-stride grid so that each voxel is covered by $2^3$ patches, each patch is multiplied by a separable 3D Hann window $W(n,m,l)=w(n)w(m)w(l)$ with $w(k)=\tfrac{1}{2}(1-\cos\tfrac{2\pi(k+\tfrac12)}{32})$, and the reconstruction is the windowed sum divided by the accumulated weight. This is the standard tiled-decode pattern of latent-diffusion decoders~\cite{rombach2022ldm,hacohen2025ltxvideo}; it is parameter-free, needs no retraining, and blends the seams that remain after training, at a one-time inference cost from the overlap.

\subsection{Training Objective}
\label{subsec:objective}
A purely pixel-wise loss is not enough: minimizing reconstruction error alone tends to smooth away the smallest scales that carry dissipation and set the spectral slope, and says nothing about the latent being well-behaved or the patch seams being consistent. We therefore train with an objective in which each term addresses a specific failure mode. An $L_1$ reconstruction term on the field drives pixel-level accuracy; a separate $L_1$ term on the spatial gradient of the field targets the small-scale, high-wavenumber structure that carries dissipation and sets the spectral slope; a Kullback--Leibler term regularizes the latent posterior toward a standard normal, keeping the latent space smooth and usable downstream; and an adversarial term~\cite{goodfellow2014gan}, supplied by a 3D patch discriminator with a hinge loss~\cite{lim2017hinge}, counteracts the over-smoothing that pixel losses induce and restores high-frequency detail.

The remaining term makes the per-patch map insensitive to where a patch boundary falls, and is the main novelty of our objective. We ask the compressor to commute with spatial translation: shifting the input and then encoding--decoding should match encoding--decoding and then shifting. Writing $S_k$ for a circular shift of $k=8$ voxels along one spatial axis, we penalise
\begin{equation}
  L_{\mathrm{consist}}
  = \big\lVert\, D_\phi(E_\theta(S_k x)) - S_k\,\mathrm{sg}\!\big[D_\phi(E_\theta(x))\big] \big\rVert_1,
  \label{eq:consist}
\end{equation}
where $\mathrm{sg}[\cdot]$ stops gradients through the unshifted branch. We fix the shift to $8$ voxels, the spatial extent of the latent, so that shifting the input by this amount corresponds to moving the latent by one grid cell. Encouraging this shift-equivariance makes the per-patch map insensitive to where a patch boundary happens to fall, which suppresses seam discontinuities and tightens the recovered physics. The full generator objective combines these five terms,
\begin{equation}
  L_G = L_{1} + \lambda_{\mathrm{grad}} L_{\mathrm{grad}} + \beta_{\mathrm{KL}} L_{\mathrm{KL}} + \lambda_{\mathrm{adv}} L_{\mathrm{adv}} + \lambda_{\mathrm{consist}} L_{\mathrm{consist}},
  \label{eq:loss}
\end{equation}
trained jointly with a 3D patch discriminator under a standard hinge objective. The loss weights are given in Section~\ref{subsec:setup_exp}.

% Section 4 — Experiments. beta_gt = -1.518 (measured fit window).
\section{Experiments}
\label{sec:experiments}

\subsection{Experimental Setup}
\label{subsec:setup_exp}
All experiments use the forced isotropic turbulence of the Johns Hopkins Turbulence Database (JHTDB)~\cite{perlman2007jhtdb,wan2016jhtdb}, an incompressible flow on a periodic $[0,2\pi]^3$ domain at Taylor-scale Reynolds number $\mathrm{Re}_\lambda \approx 418$, with the three velocity components and pressure as the four channels. We train at $256^3$ and evaluate at $1024^3$. The training fields are obtained from the native $1024^3$ data by stride-four downsampling, so that no model ever sees a $1024^3$ field during training; the $1024^3$ test frames are therefore a genuine zero-shot, higher-resolution target rather than a held-out split of the training distribution. We evaluate on three effectively uncorrelated test frames $\{800,900,1000\}$ and report their mean. Fields are normalized to $[-1,1]$ (so PSNR uses peak $2.0$), and all reconstruction times are measured on a single RTX~5090. JHTDB stores the flow in dimensionless simulation units (viscosity $\nu=1.85\times10^{-4}$, domain side $2\pi$), so we report physical quantities as dimensionless values or as ratios of reconstruction to DNS ground truth; the only dimensioned quantity is PSNR, in decibels.

The compressor is trained with AdamW~\cite{loshchilov2019adamw} in mixed precision, using a generator learning rate of $10^{-4}$ and a slower discriminator rate of $5\times 10^{-5}$ to keep the adversarial game balanced, with gradient clipping and a cosine schedule. In the generator objective of Eq.~\eqref{eq:loss} we set the loss weights to $\lambda_{\mathrm{grad}}{=}0.5$, $\beta_{\mathrm{KL}}{=}0.01$, $\lambda_{\mathrm{adv}}{=}0.01$, and $\lambda_{\mathrm{consist}}{=}0.1$, with a consistency shift of $8$ voxels; these were selected by an ablation grid on the $256^3$ training set before any $1024^3$ evaluation frame was touched. The generator and discriminator together hold roughly $45$M parameters, and training converges quickly from a warm start. The learned baselines use the configurations recommended in their original papers, without further tuning, and the analytic methods have no learnable parameters: their only knob is the compression ratio (e.g.\ POD modes, TT-SVD bond dimension, ZFP accuracy), fixed to the $64\times$ tier. Each learned model is trained once; the analytic methods are deterministic, so the variation we report (Figure~\ref{fig:variation}) is across the three held-out $1024^3$ evaluation frames of a single trained model rather than across retrainings.

At test time every method runs through an identical per-frame pipeline so that all numbers are comparable: each $1024^3$ field is normalized, reconstructed, mapped back to physical units, and scored, with physics quantities computed spectrally on the GPU. Only the reconstruction call itself is timed, on the same single consumer GPU, so the reported times are directly comparable.

\subsection{Baselines}
\label{subsec:baselines}
We compare against three families of compressors, all evaluated at the same $64\times$ tier (about $2{,}048$ stored values per $32^3$ patch) so that the comparison is like-for-like. The first family is classical per-frame methods that fit each field directly and use no learning: proper orthogonal decomposition (POD)~\cite{lumley1967pod,sirovich1987snapshot}, Daubechies wavelet thresholding~\cite{daubechies1988wavelets}, the fixed-rate floating-point compressor ZFP~\cite{lindstrom2014zfp}, and a quantics tensor-train (TT) decomposition~\cite{oseledets2011tt,schollwock2011dmrg}. We also include the most trivial $64\times$ compressor possible, a $4^3$ box-average followed by trilinear upsampling (\emph{Stride-4}), as a no-method floor: it uses no learning and no fitting, and a learned compressor should clearly beat it to justify its training, since otherwise a plain low-pass filter would suffice at this ratio. The second family is learned autoencoders trained directly on $1024^3$ patches, covering the dominant generative-compression backbones, namely the Stable Diffusion VAE (SD-VAE)~\cite{rombach2022ldm}, the Deep Compression Autoencoder (DC-AE)~\cite{chen2025dcae}, and the Representation Autoencoder (RAE)~\cite{zheng2025rae}, together with three 2025 state-of-the-art video tokenizers ported to 3D: the Wavelet Flow VAE (WF-VAE)~\cite{li2025wfvae}, the Cosmos tokenizer~\cite{nvidia2025cosmos}, and the LTX-Video VAE~\cite{hacohen2025ltxvideo}. The third family is the zero-shot setting that our method targets: trained at $256^3$ and applied unchanged at $1024^3$.

Two of the learned baselines are reported with naive non-overlapping reassembly rather than the Hann overlap-add used for the other learned methods, because the overlap-add is incompatible with them: for Cosmos the blended reconstruction diverges, and for LTX-Video it degrades the spectral slope error $|\Delta\beta|$ by more than twenty-fold. We therefore report their stronger naive results and mark them accordingly.

A per-frame conditional neural field (CNF)~\cite{guo2025cnf} and a function-space autoencoder (FAE)~\cite{bunker2025fae} are conceptually related but not tier-comparable, as CNF fits a per-frame decoder and FAE stores a single global latent rather than a spatial-tensor latent; we therefore exclude them from the main comparison.

\subsection{Metrics}
\label{subsec:metrics}
Because our central claim is about physics rather than appearance, the primary metrics are physical. We report the energy-dissipation rate and the enstrophy as ratios of reconstruction to ground truth, both of which should equal one when the physics is preserved. We characterize the energy spectrum by its inertial-range slope, fit over wavenumbers $k\in[4,40]$, and report the absolute slope error $|\Delta\beta|=|\beta_{\mathrm{recon}}-\beta_{\mathrm{gt}}|$ between the reconstructed and ground-truth slopes. The measured $\beta_{\mathrm{gt}}\approx-1.52$ is slightly shallower than the theoretical $-5/3$, reflecting the limited inertial range at $\mathrm{Re}_\lambda\approx418$: the $[4,40]$ fit window includes part of the upper energy-containing range (whose peak lies near $k\approx2$--$4$), which biases the global fit toward a shallower slope. We therefore evaluate the reconstructed slope against the measured $\beta_{\mathrm{gt}}$ rather than against $-5/3$. We also report the relative divergence $\lVert\nabla\!\cdot\!u\rVert_2/\mathrm{std}(u)$, which probes how well the incompressibility constraint is respected. Reconstruction quality is summarized by the relative $L_1$ and $L_2$ errors, RMSE, and PSNR. Finally, we report model size as parameter count.

\subsection{Main Results}
\label{subsec:main_results}
% Table 1 — main comparison @ 64x: reconstruction + physics + time in one table*.
\begin{table*}[tbp]
  \centering
  \caption{\textbf{Main comparison at the $64\times$ tier} on zero-shot $1024^3$
  frames (metrics in Section~\ref{subsec:metrics}; $\varepsilon,\Omega$ target $1$;
  Time is the single-pass per-frame reconstruction time in seconds).
  \textbf{Bold}/\underline{underline}: best/second per column. The
  \colorbox{controlrow}{shaded} row is our in-distribution upper bound, trained at
  $1024^3$ under the same protocol as the learned baselines.}
  \label{tab:main}
  \footnotesize
  \setlength{\tabcolsep}{8pt}
  \begin{tabular}{l rrrr rrrr r}
    \toprule
    & \multicolumn{4}{c}{\textit{Reconstruction}} & \multicolumn{4}{c}{\textit{Physics fidelity}} & \\
    \cmidrule(lr){2-5}\cmidrule(lr){6-9}
    Method & rel-$L_1$\,\down & rel-$L_2$\,\down & RMSE\,\down & PSNR\,\up & $\varepsilon$\,\toone & $\Omega$\,\toone & $|\Delta\beta|$\,\down & $\nabla\!\cdot\!u$\,\down & Time\,\down \\
    \midrule
    \multicolumn{10}{@{}l}{\textit{Per-frame analytic}} \\
    Stride-4$^{\sharp}$ & 0.076 & 0.090 & 0.061 & 27.04 & 0.369 & 0.362 & 0.104 & \textbf{2.99} & \textbf{18} \\
    POD & 0.056 & 0.063 & 0.042 & 30.19 & 1.205 & 1.153 & 0.005 & 7.75 & 160 \\
    Wavelet (db4, level 3) & 0.050 & 0.052 & 0.035 & 31.69 & 0.869 & 0.828 & \textbf{0.001} & 6.99 & 664 \\
    ZFP & 0.050 & 0.062 & 0.042 & 30.25 & 2.019 & 1.684 & 0.002 & 19.76 & 151 \\
    TT-SVD / MPS (bond=9) & 0.118 & 0.131 & 0.088 & 23.83 & 1.064 & 0.879 & $0.362^{\ddagger}$ & 15.94$^{\ddagger}$ & 427 \\
    \midrule
    \multicolumn{10}{@{}l}{\textit{ML trained at $1024^3$}} \\
    SD-VAE-3D & 0.056 & 0.061 & 0.041 & 30.42 & \underline{0.979} & 0.957 & 0.029 & 6.02 & 94 \\
    DC-AE-3D & 0.044 & 0.050 & 0.034 & 32.08 & 0.824 & 0.806 & 0.002 & 5.75 & 147 \\
    RAE-3D & 0.104 & 0.108 & 0.073 & 25.42 & 0.400 & 0.393 & 0.128 & 5.92 & 80 \\
    WF-VAE-3D & 0.036 & 0.042 & 0.028 & 33.87 & 0.902 & 0.888 & 0.002 & 5.45 & 95 \\
    Cosmos-CV-3D$^{\flat}$ & 0.062 & 0.071 & 0.048 & 29.07 & 1.333 & 1.191 & 0.005 & 13.05 & 112 \\
    LTX-Video-VAE-3D$^{\flat}$ & 0.078 & 0.090 & 0.061 & 27.14 & 1.519 & 1.376 & \textbf{0.001} & 12.93 & 474 \\
    \midrule
    \multicolumn{10}{@{}l}{\textit{Ours}} \\
    \textbf{PPLC (zero-shot, $256^3\!\to\!1024^3$)} & \underline{0.035} & \underline{0.040} & \underline{0.027} & \underline{34.17} & 1.056 & \textbf{1.023} & 0.005 & 6.23 & \underline{47} \\
    \rowcolor{controlrow}
    \textit{PPLC (trained at $1024^3$)} & \textbf{0.032} & \textbf{0.039} & \textbf{0.026} & \textbf{34.39} & \textbf{1.016} & \underline{0.965} & \textbf{0.001} & \underline{5.04} & \underline{47} \\
    \bottomrule
  \end{tabular}

  \vspace{2pt}
  {\footnotesize Learned rows use the same Hann overlap-add reassembly; analytic rows use their native global transform.
  $^{\sharp}$Stride-4's low divergence is an artifact of trilinear upsampling
  (smooth by construction), not fidelity. $^{\ddagger}$TT-SVD's $|\Delta\beta|$ and
  $\nabla\!\cdot\!u$ are from a legacy spectral operator; $\varepsilon,\Omega$
  unaffected. $^{\flat}$Cosmos and LTX-Video are reported with naive reassembly:
  overlap-add diverges for Cosmos and badly degrades LTX-Video's slope error and speed.}
\end{table*}

\begin{figure*}[t]
  \centering
  \includegraphics[width=\textwidth]{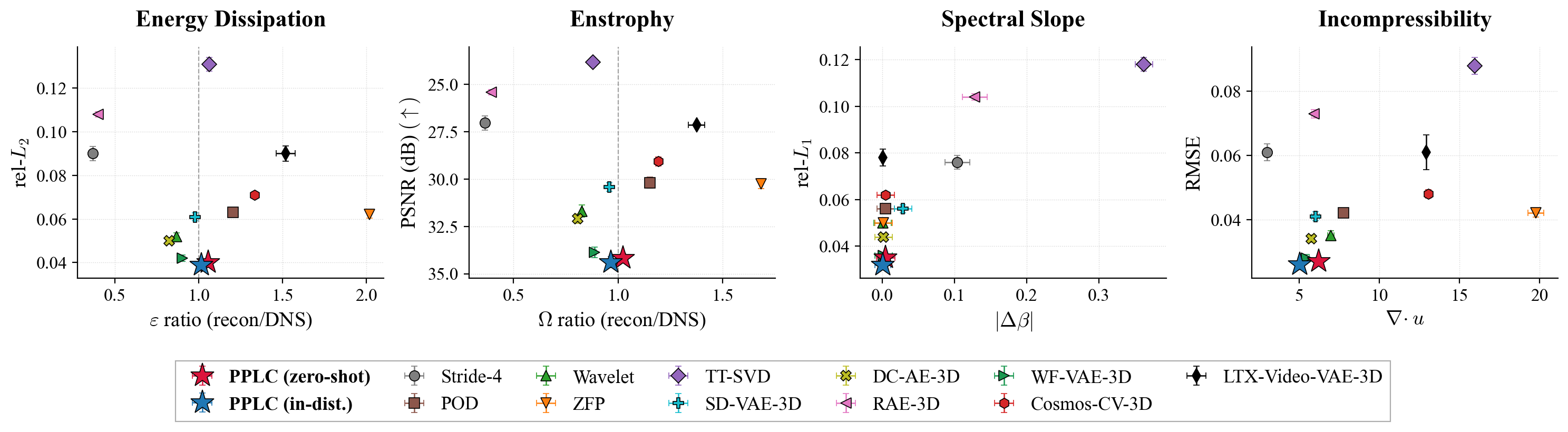}
  \caption{\textbf{Mean and per-frame variation across the three $1024^3$ test
  frames} behind Table~\ref{tab:main}. Each panel pairs a reconstruction metric
  with a physics metric, with error bars (one standard deviation over the three
  frames) on both axes. The $y$ axis is always better lower (the PSNR axis of the
  second panel is inverted accordingly); on the $x$ axis, the first two panels are
  best near the dashed unity line ($\varepsilon,\Omega\!\to\!1$) and the last two
  toward the origin. Stars mark our two PPLC variants.}
  \label{fig:variation}
\end{figure*}

\begin{figure*}[t]
  \centering
  \includegraphics[width=\textwidth]{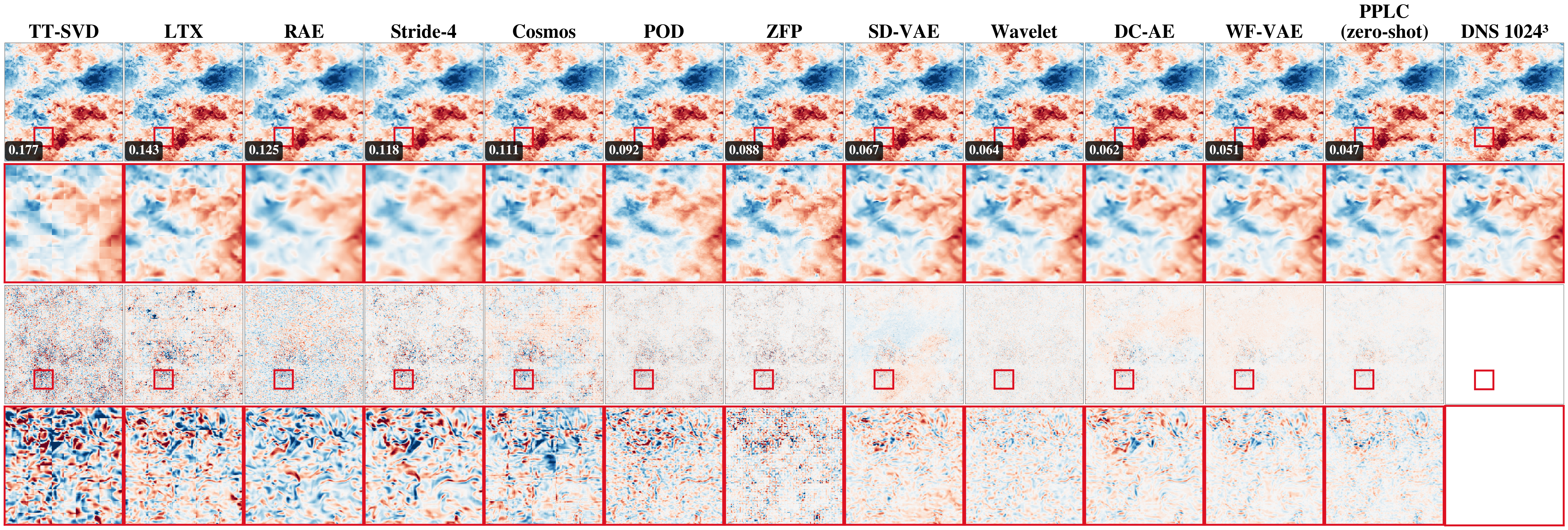}
  \caption{\textbf{Qualitative comparison on a zero-shot $1024^3$ slice.} Methods
  are ordered left to right by decreasing per-slice rel-$L_2$ (top-row labels),
  ending with PPLC and the DNS ground truth. Rows, top to bottom: reconstruction
  (red box = zoom region), the zoom, the error map, and the zoomed error map.}
  \label{fig:qualitative}
\end{figure*}

\begin{figure}[t]
  \centering
  \includegraphics[width=\columnwidth]{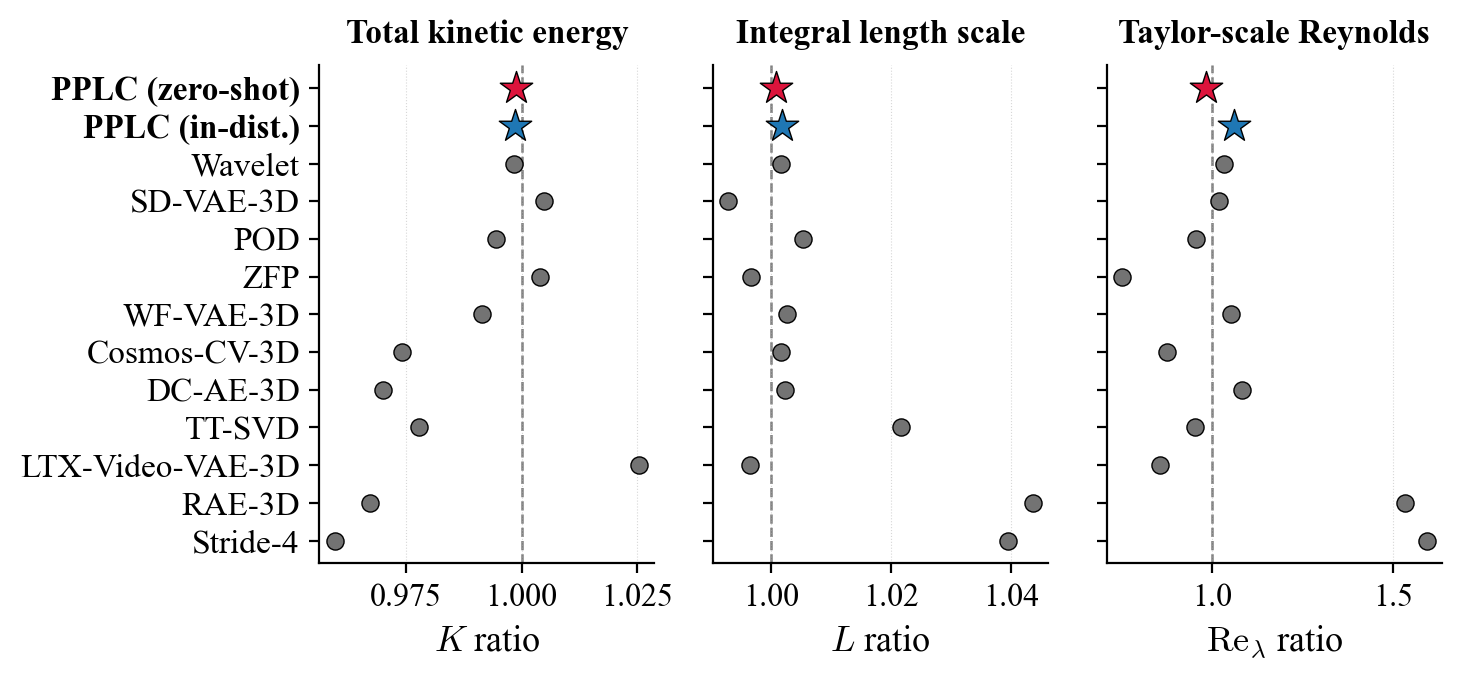}
  \caption{\textbf{Large-scale integral quantities} on the zero-shot $1024^3$
  frame, as ratios of reconstruction to DNS (dashed line at the ideal value $1$;
  closer is better). Stars mark our two PPLC variants.}
  \label{fig:largescale}
\end{figure}

Table~\ref{tab:main} reports reconstruction quality, physics fidelity (our main concern), and reconstruction time together, all at the $64\times$ tier on the zero-shot $1024^3$ frames. Despite never seeing a $1024^3$ field during training, PPLC attains the lowest reconstruction error and highest PSNR of any method except its own variant trained at the test resolution (rel-$L_2$ $0.040$ and $34.17$\,dB, against $0.039$ and $34.39$\,dB for that in-distribution model, the shaded row, which serves as an upper-bound reference rather than a competitor). It is ahead of the strongest learned baseline, WF-VAE ($0.042$), and the best analytic compressor, Wavelet thresholding ($0.052$); the ranking is stable across rel-$L_1$, rel-$L_2$, and RMSE, so it is not an artifact of the $L_2$ metric, which for turbulent fields can over-weight the energetic large scales~\cite{zhao2017loss,wang2022l2physicsinformedlosssuitable}. On physics fidelity the picture is favorable but more nuanced. PPLC has the enstrophy ratio closest to unity of any method ($\Omega=1.023$) and the second-closest dissipation ratio ($\varepsilon=1.056$, behind only SD-VAE at $0.979$); the other learned baselines after the same reassembly either under-predict both quantities (e.g.\ WF-VAE $\varepsilon=0.902$, DC-AE $0.824$) or strongly over-predict them (Cosmos and LTX-Video $\varepsilon>1.3$), whereas PPLC keeps both near unity. Its spectral-slope error ($|\Delta\beta|=0.005$) and divergence ($6.23$) are mid-pack, but no competing method is simultaneously strong on reconstruction and on the conserved quantities, which is the combination that matters for a latent meant to feed a downstream simulator. The advantage is clearest on the large-scale integral quantities turbulence theory cares about most, the total kinetic energy $K$, integral length scale $L$, and Taylor-scale Reynolds number $\mathrm{Re}_\lambda$ (Figure~\ref{fig:largescale}): PPLC ranks first on all three (ratios to DNS $0.999$, $1.001$, $0.982$), even though every baseline is fitted or trained natively at $1024^3$ while PPLC is applied zero-shot from $256^3$; our in-distribution variant ranks a close second or third, indicating the architecture itself keeps these large scales close to the reference. The picture is thus complementary across scales: methods like SD-VAE are strongest on small-scale dissipation, whereas PPLC is consistently strongest on the large-scale, integral quantities. The per-frame variation behind these means (Figure~\ref{fig:variation}) is small relative to the gaps between methods, with both PPLC variants in the favorable region of every panel, so the ordering is stable. Figure~\ref{fig:qualitative} shows the same picture qualitatively: PPLC's zoomed reconstruction and error map are visibly the cleanest among the compressors, while the weaker baselines either over-smooth the fine structure or leave a denser error field.

The analytic baselines are informative reference points rather than competitors: POD and ZFP reconstruct reasonably but distort dissipation (ZFP doubles it), and Wavelet thresholding attains a low reconstruction error and a faithful spectral slope ($|\Delta\beta|=0.001$). Wavelet, however, produces a sparse coefficient set rather than a spatial-tensor latent, so it cannot serve as input to the downstream latent-space models that motivate compression here. Among the learned baselines the behavior is mixed: the same overlap-add reassembly that helps PPLC leaves several of them under-predicting dissipation and enstrophy (RAE collapses to $\varepsilon=0.400$), so no single baseline is strong on reconstruction and physics at once. The headline result is therefore not any single number but the setting: PPLC matches or exceeds compressors trained directly at $1024^3$, while its own training data is a stride-downsampled $256^3$ field that is far cheaper to produce.

\begin{figure}[t]
  \centering
  \includegraphics[width=\columnwidth]{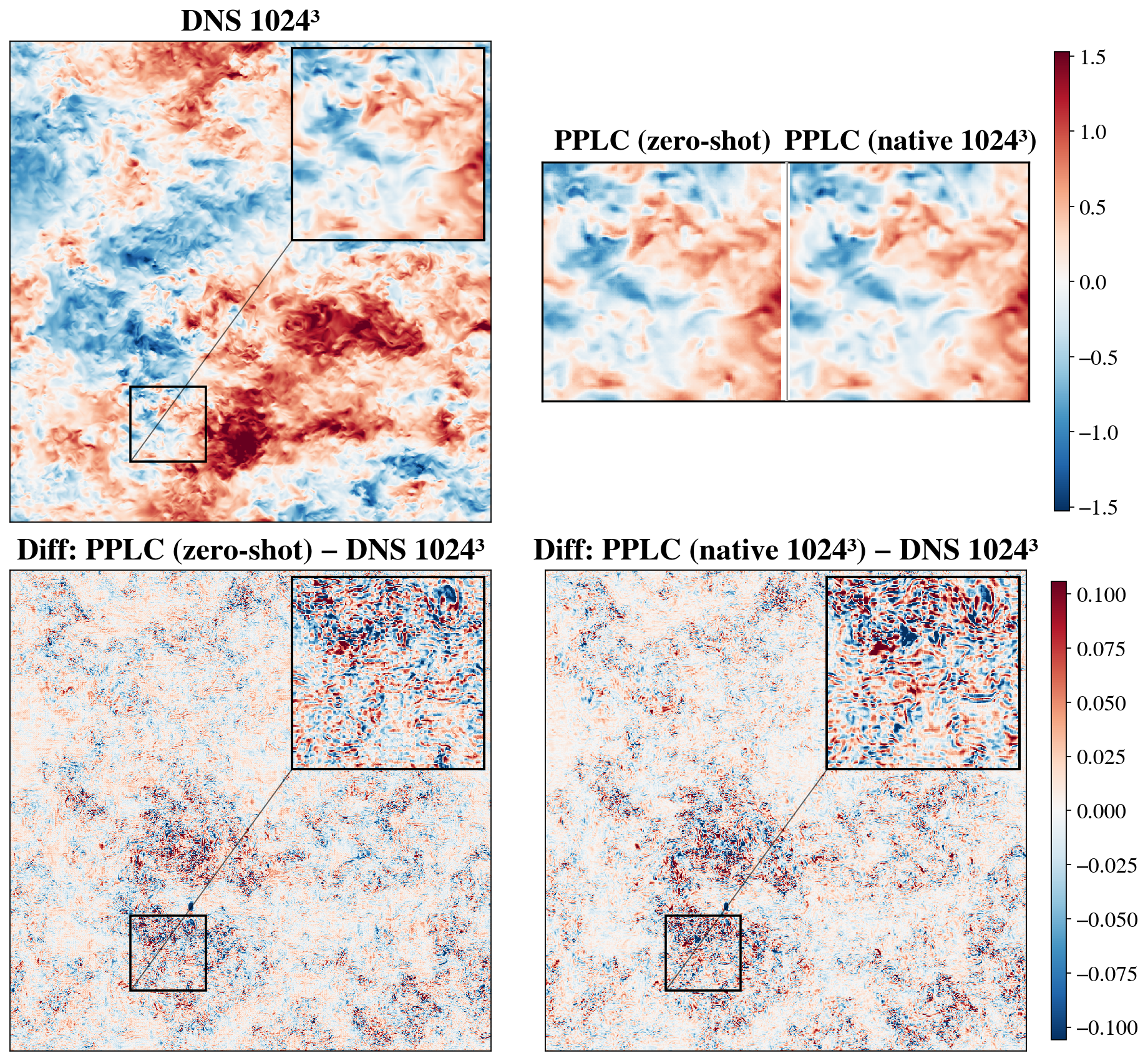}
  \caption{\textbf{Zero-shot vs.\ in-distribution} (PPLC) on a $1024^3$ slice.
  Top: DNS ground truth and the two reconstructions; bottom: their error maps
  with a magnified inset. Each row shares a colorbar.}
  \label{fig:zeroshot_vs_native}
\end{figure}

This variant, \emph{PPLC (in-dist., $1024^3$)} (the shaded row), is identical to the zero-shot model in architecture, hyperparameters, optimizer, and early-stopping patience, and differs only in its training data: native $1024^3$ patches rather than stride-four $256^3$ fields. The gap between the two therefore measures exactly what zero-shot transfer costs. On reconstruction it is almost nil: the zero-shot model reaches rel-$L_2$ $0.040$ against the control's $0.039$ and is within $0.22$\,dB on PSNR, so seeing the test resolution during training buys essentially no reconstruction advantage. The physics is likewise close. Both keep the spectral slope within $0.4\%$ of ground truth ($\beta_{\mathrm{recon}}=-1.523$ zero-shot versus $-1.517$ in-distribution, against $\beta_{\mathrm{gt}}=-1.518$), and both recover dissipation and enstrophy to within a few percent ($\varepsilon=1.056$, $\Omega=1.023$ zero-shot versus $1.016$, $0.965$ in-distribution), each slightly over- or under-shooting by a comparable margin. The practical reading is that the $4\times$ resolution mismatch costs almost nothing, on either reconstruction or physics, which we take as evidence that zero-shot cross-resolution transfer is a viable strategy when training data at the test resolution is unavailable. Figure~\ref{fig:zeroshot_vs_native} illustrates this on a single slice: the two reconstructions have very similar slice errors ($0.0475$ and $0.0490$) and differ mainly in the texture of the residual, not its magnitude, which we read as essentially on par.

The same setting yields two practical advantages. First, inference is fast: a full $1024^3$ field is reconstructed in $47$\,s on a single consumer GPU (Table~\ref{tab:main}, single-pass reconstruction time), about $2\times$ faster than the strongest learned baseline (WF-VAE, $95$\,s), because the small spatial latent and factorized convolutions make patch reconstruction trivially parallel with no inter-patch attention. Second, data acquisition is far cheaper: a single four-channel $1024^3$ frame in single precision occupies $16$\,GB, whereas its stride-four $256^3$ counterpart is $256$\,MB, exactly $64\times$ smaller; across our training set this is the difference between $251$\,GB and $15.7$\,TB. This saving is in data handling, not in the underlying simulation, since both paths reuse the same archived DNS.

\subsection{Ablations}
\label{subsec:ablations}
\begin{figure}[t]
  \centering
  \includegraphics[width=\columnwidth]{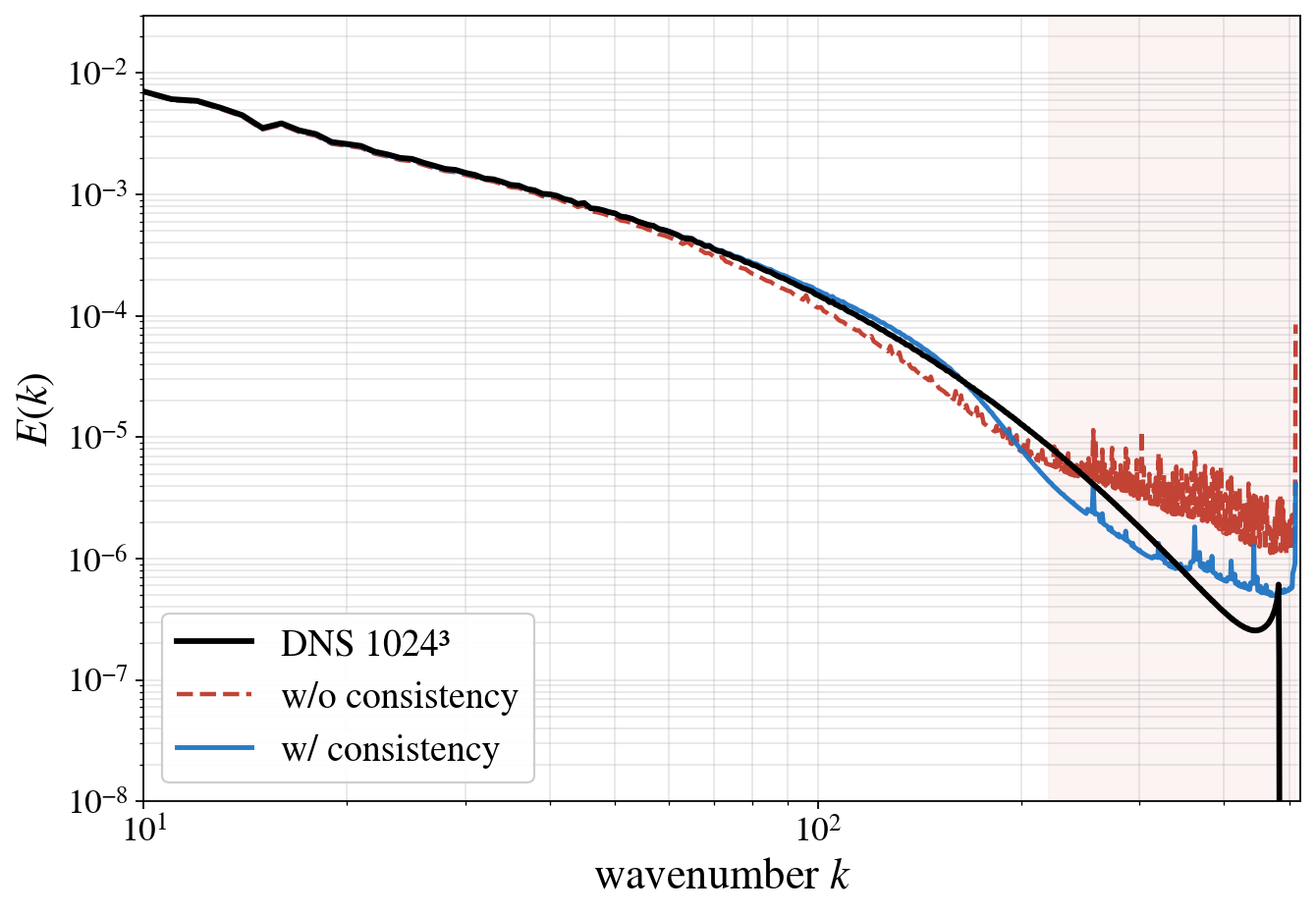}
  \caption{\textbf{Effect of the consistency loss on the energy spectrum.}
  Without the loss the high-wavenumber tail (shaded) oscillates strongly; with it
  the spectrum stays smooth and close to the DNS reference, improving the
  small-scale content at little cost to the aggregate metrics.}
  \label{fig:spectrum}
\end{figure}

We isolate the contribution of three design choices, the latent layout, the consistency loss, and the overlap-add reassembly, and study how the method behaves as the compression ratio is pushed. Table~\ref{tab:ablation} builds the model up one component at a time at a fixed $64\times$ budget. A channel-heavy $32\times4^3$ latent gives the weakest result (rel-$L_2$ $0.076$, divergence $14.97$), and adding the consistency loss to it helps the physics only modestly ($\varepsilon$ $1.32\!\to\!1.20$) because the divergence is dominated by the latent shape, not the seam: switching to our spatial-layout $4\times8^3$ latent is what brings the divergence down sharply (to $9.75$ once the consistency loss is also on) while improving reconstruction to $0.052$. Replacing the naive non-overlapping tiling with the Hann overlap-add then improves every metric again (rel-$L_2$ $0.052\!\to\!0.040$, $\varepsilon$ $1.19\!\to\!1.06$, divergence $9.75\!\to\!6.23$), confirming that the residual error of the trained model is concentrated at the patch boundaries. The spatial latent, the consistency loss, and the overlap-add are thus complementary: the layout preserves neighborhood structure, the loss attacks the seam at training time, and the overlap-add blends what remains at test time, and only with all three does the dissipation ratio come close to unity. The consistency loss has an effect that the aggregate metrics understate, visible in the energy spectrum (Figure~\ref{fig:spectrum}): without it, the reconstructed spectrum develops large spurious oscillations at high wavenumbers, whereas with it the high-$k$ tail stays smooth and close to the DNS reference. This region contributes little to a pixel-space error, so the gain in Table~\ref{tab:ablation} is small, but it is exactly the small-scale content a downstream model would consume, so the cleaner high-frequency latent matters well beyond what the reconstruction numbers suggest.

 Table~\ref{tab:tiers} applies the same recipe across compression tiers from $34\times$ to $254\times$; to isolate the ratio effect it uses naive reassembly throughout, so its rows sit slightly above the Hann headline. The method degrades gracefully along a clean quality-versus-ratio curve: $64\times$ is the near-lossless operating point we report as headline, $128\times$ is the practical ceiling (physics still valid, with the slope beginning to drift), and $254\times$ shows mild but real degradation of dissipation and slope.

We also examine the role of training resolution. Training at $128^3$ instead of $256^3$ noticeably degrades both reconstruction and dissipation at the $64\times$ tier, and at the more aggressive $128\times$ tier the recovered dissipation departs sharply from ideal, because a $128^3$ grid truncates the high-wavenumber structure the encoder needs to learn; $256^3$ is thus the practical floor for zero-shot transfer to $1024^3$.
% Table 2 — component ablation @ 64x (one independent float).
\begin{table}[t]
  \centering
  \caption{\textbf{Component ablation} at $64\times$, added one at a time (last
  row is our full model).}
  \label{tab:ablation}
  \small
  \begin{tabular}{@{}l c c c c c@{}}
    \toprule
    Variant & latent & $\lambda_{\mathrm{c}}$ & rel-$L_2$\,\down & $\varepsilon$\,\toone & $\nabla\!\cdot\!u$\,\down \\
    \midrule
    channel-heavy            & $32{\times}4^3$ & 0   & 0.076 & 1.32 & 14.97 \\
    channel-heavy, +consist  & $32{\times}4^3$ & 0.1 & 0.074 & 1.20 & 14.65 \\
    spatial-layout           & $4{\times}8^3$  & 0   & 0.062 & 1.27 & 17.6 \\
    spatial, +consist (naive) & $4{\times}8^3$ & 0.1 & 0.052 & 1.19 & 9.75 \\
    \textbf{\quad + Hann (ours)} & $\mathbf{4{\times}8^3}$ & \textbf{0.1} & \textbf{0.040} & \textbf{1.06} & \textbf{6.23} \\
    \bottomrule
  \end{tabular}
\end{table}

% Table 3 — compression-tier sweep (naive reassembly), one independent float.
\begin{table}[t]
  \centering
  \caption{\textbf{Compression-tier sweep} (naive reassembly, to isolate the
  ratio effect).}
  \label{tab:tiers}
  \small
  \begin{tabular}{@{}l c c c c c@{}}
    \toprule
    Tier & latent & rel-$L_2$\,\down & PSNR\,\up & $\varepsilon$\,\toone & $\beta_{\mathrm{recon}}$ \\
    \midrule
    $34\times$           & $60{\times}4^3$ & 0.033 & 35.27 & 1.07 & $-1.530$ \\
    $64\times$  & $4{\times}8^3$  & 0.052 & 31.80 & 1.191 & $-1.522$ \\
    $128\times$          & $2{\times}8^3$  & 0.083 & 28.74 & 1.53 & $-1.541$ \\
    $254\times$          & $1{\times}8^3$  & 0.141 & 25.34 & 1.99 & $-1.579$ \\
    \bottomrule
  \end{tabular}
\end{table}

\subsection{Preliminary: Forecasting in the Latent}
\label{subsec:downstream}
The compressor above is our main contribution; here we add a small experiment to check whether its latent can also serve as a state space for simulation, not only for reconstruction. It is deliberately separate from the zero-shot setting, run entirely \emph{in distribution} at $256^3$ with the same $64\times$ compressor, and is meant only as a sanity check rather than a full forecasting study. We train forecasters that, given a short history of context frames, predict the turbulent field $\tau$ simulation steps ahead, and compare forecasting directly in pixel space against forecasting in the compressed latent (predicting in the latent and decoding before scoring, so both are evaluated in pixel space). As backbones we use a Transformer and a UNet, the two architectures behind the most representative vanilla models for JHTDB turbulence simulation~\cite{dai2026pest,dai2026flowrefiner}, and we consider the two classical rollout strategies: direct query of a fixed horizon and autoregressive (AR) stepping.

\noindent\textbf{Setup.} Given a snapshot at time $t$, the task is to predict the field $\tau$ simulation steps ahead (each step $\Delta t{=}0.01$), where a forecaster either queries the target horizon directly or reaches it autoregressively. All forecasters operate in the latent of a \emph{frozen} $64\times$ PPLC encoder, on the $256^3$ JHTDB split (frames $0$--$800$ train, $800$--$900$ validation, $900$--$1000$ test), scored by $\mathrm{FSS}_\tau$ against a persistence baseline. The best model (latent Transformer, direct-$\tau$) regresses $z_{t+\tau}$ from a three-frame context $\{z_{t-10},z_{t-5},z_t\}$ for a sampled $\tau$, and is trained with conditional image leakage~\cite{zhao2024cil} ($\beta_m{=}0.5$), a $\tau$-dependent noise schedule on the context that mimics the upstream errors of a real rollout and prevents over-fitting to clean, oracle context; its single-shot prediction avoids error accumulation. The UNet variant instead fixes $\tau{=}10$ and reaches longer horizons autoregressively; we use a single-step jump of $10$ rather than a smaller one because, given the strong short-time correlation of the flow, we found small steps to predict little beyond the input and to compound error quickly under rollout.

We report rollout RMSE at $\tau{=}10$ and $\tau{=}20$, but since pixel-space error is dominated by how much the flow naturally decorrelates over $\tau$ steps, raw RMSE is hard to compare across horizons; our headline metric is therefore the Forecast Skill Score (FSS)~\cite{murphy1988skill}, the standard forecast-verification metric that normalizes a model's error by a reference baseline, at $\tau{=}20$ (about $4.5$ Kolmogorov times). We define it as $1-\lVert\hat u_\tau-u_\tau\rVert/\lVert u_{t_0}-u_\tau\rVert$, the relative error reduction over an identity baseline that simply assumes the flow does not change. A positive FSS means the forecaster extracts genuinely useful dynamics; a negative FSS means it does worse than predicting no change at all. As Table~\ref{tab:forecaster} shows, only the direct-$\tau$ Transformers operating in our latent achieve positive FSS, the best reaching $+0.070$, while the pixel-space models and the autoregressive UNet baselines all fall below zero. Two controlled comparisons isolate why. Holding the backbone fixed, the same Transformer reaches $+0.070$ in our latent but only $-0.110$ in pixel space, so the gain comes from the compressed state space rather than the predictor. Part of the difficulty in pixel space is that a full $256^3$ field does not fit a forecaster directly: the pixel UNet has to tile the volume into octants and the pixel Transformer has to tokenize it into large patches, both of which fragment the global structure that our compact latent instead presents as a whole. Holding the domain fixed, the latent UNet rolled out autoregressively drops to $-0.064$, indicating that a direct-$\tau$ predictor avoids the error accumulation of autoregressive rollout. This shows the latent is not merely decodable but a usable state space for simulation. We evaluate this in distribution at $256^3$; coupling it with the zero-shot reconstruction setting to forecast at $1024^3$ is a natural next step.
% Table — preliminary forecasting (256^3 in-distribution): RMSE@tau=10,20 + FSS@tau=20.
\begin{table}[t]
  \centering
  \caption{\textbf{Preliminary forecasting results} ($256^3$ in-distribution).}
  \label{tab:forecaster}
  \small
  \begin{tabular}{@{}l l l c c c@{}}
    \toprule
    & & & \multicolumn{2}{c}{RMSE\,\down} & FSS\,\up \\
    \cmidrule(lr){4-5}
    Domain & Backbone & Rollout & $\tau{=}10$ & $\tau{=}20$ & $\tau{=}20$ \\
    \midrule
    \textbf{latent} & \textbf{Transformer} & \textbf{direct} & \textbf{0.273} & \textbf{0.332} & $\mathbf{+0.070}$ \\
    latent & UNet ($\tau{=}10$) & AR & 0.303 & 0.423 & $-0.064$ \\
    pixel & Transformer & direct & 0.296 & 0.452 & $-0.110$ \\
    pixel & UNet ($\tau{=}10$) & AR & 0.178 & 0.474 & $-0.141$ \\
    \bottomrule
  \end{tabular}

  \vspace{2pt}
  {\footnotesize Pixel rows operate on the $256^3$ JHTDB data directly; latent
  rows forecast in our compressed latent and decode before scoring, so both are
  evaluated in pixel space. UNet rows are trained at $\tau{=}10$ and reach
  $\tau{=}20$ autoregressively (AR), which accumulates error.}
\end{table}

% ============================================================================
% Section 5 — Conclusion  (~0.3 page). PROSE. DRAFTED 2026-06.
% Recap -> why it works -> downstream -> limitations/future. No new content,
% no result numbers (those live in Sec.4). Terminology unified (simulation),
% no em-dashes, no overclaim.
% ============================================================================
\section{Conclusion}
\label{sec:conclusion}

This paper proposed \emph{\textbf{Physics-Preserving Latent Compression (PPLC)}} for high-ratio compression and zero-shot resolution transfer in three-dimensional turbulence. The key idea is to use inertial-range scale similarity to make local patches transferable across resolutions, implemented with a shared patch-local variational autoencoder that applies unchanged from low- to high-resolution fields. Experiments on forced isotropic turbulence show that PPLC transfers zero-shot from stride-downsampled $256^3$ to $1024^3$ fields, balancing compression efficiency, reconstruction accuracy, and physical fidelity. Because compression is lossy, no method recovers the physics exactly; the reconstructed fields nonetheless keep the key physical diagnostics closer to the ground truth than the baselines, and a preliminary forecasting experiment suggests the learned latent can support downstream turbulence dynamics. As a general-purpose framework for physics-preserving latent compression, PPLC holds promise beyond isotropic turbulence, paving the way for scalable surrogate modeling across scientific domains.

% No \section*{Acknowledgment} — removed for triple-blind review.

% Flush all queued table floats before the bibliography so tables do not
% interleave with the reference list. \clearpage outputs every queued float,
% then starts the references on a fresh page.
\clearpage

% ----------------------------------------------------------------------------
% REFERENCES — references.bib
% ----------------------------------------------------------------------------
\bibliographystyle{IEEEtran}
\bibliography{references}

% ----------------------------------------------------------------------------
% Anonymization reminder for final PDF export:
%   strip PDF metadata, e.g. \usepackage[pdfauthor={}]{hyperref} or post-process.
% ----------------------------------------------------------------------------

\end{document}